\documentclass{aa}  

\usepackage{graphicx}
\usepackage{txfonts}
\usepackage{lipsum}
\usepackage{subcaption}    
\usepackage{lscape}
\usepackage{placeins} 
\usepackage{natbib}
\bibpunct{(}{)}{;}{a}{}{,}
                                
\usepackage{hyperref}
\begin{document}

   \title{Determining the incidence rate of magnetic $\delta$~Scuti candidates with \textit{CoRoT}}
   \author{G. Paul\inst{\ref{inst1}}
        \and
        C. Neiner\inst{\ref{inst1}}
        \and
        C. Catala\inst{\ref{inst1}}
        \and
        J. Labadie-Bartz\inst{\ref{inst2},\ref{inst1}}
          }

   \institute{LIRA, Paris Observatory, CNRS, PSL University, Sorbonne University, Universit\'e Paris Cit\'e, CY Cergy Paris University, 5 place Jules Janssen, 92195 Meudon, France\\ \email{gautam.paul@obspm.fr} \label{inst1}
        \and
        DTU Space, Technical University of Denmark, Elektrovej 327, Kgs., Lyngby 2800, Denmark\label{inst2}
        }
   \date{1 November 2025}

  \abstract
    {$\delta$~Scuti stars are pulsating stars constituting the $\delta$~Scuti instability strip in the Hertzsprung-Russell (HR) diagram, which consists of A and F stars of various evolutionary stages. They are in the transition region between high-mass hot stars and low-mass solar-like stars, making understanding their magnetic properties essential to painting a complete picture of magnetism across the HR diagram. Furthermore, discovering magnetic $\delta$~Scuti stars allows for magneto-asteroseismology, which can be used to determine the internal rotation profile, internal magnetic field strength, and the efficiency of mixing and transport processes more accurately than classical asteroseismology.}
    {To date, magnetic fields have been detected at the surface of 13 $\delta$~Scuti stars. However, the overall incidence rate of magnetism in these stars remains unknown. Fossil magnetic fields are detected in 10\% of OBA stars. The goal of this work is to find out if it is the same for $\delta$~Scuti stars.}
    {We investigated the incidence rate of surface magnetic fields among $\delta$ Scuti stars using photometric data from the CoRoT space mission. We analyzed long-duration ($\sim$ 5 months) light curves of $\sim$1750 $\delta$~Scuti stars to search for pulsations and rotational modulation -- a photometric signature that indicates chemical or temperature spots at the stellar surface, usually caused by magnetic fields.}
    {We identified 147 rotational variables that we designate as magnetic candidates, thus potentially increasing the known population of magnetic $\delta$ Scuti stars drastically and suggesting an incidence rate of fossil magnetic fields in $\delta$ Scuti stars similar to the incidence rate in OBA stars in general. Our analysis also revealed a few $\delta$~Scuti--$\gamma$~Dor hybrid stars and four binary stars in the sample. We determined the rotation periods and projected rotation velocities of the magnetic candidates in order to select suitable targets for follow-up spectropolarimetric observations aimed at confirming and characterizing their magnetic fields. 
    }
    {}

   \maketitle
  \nolinenumbers

\section{Introduction}

The intermediate mass (1.5 $M_\odot$ to 2.5 $M_\odot$, A-F spectral type) range on the main sequence of the Hertzsprung-Russell (HR) diagram includes $\gamma$~Doradus and $\delta$~Scuti variable stars.
$\gamma$~Doradus stars host low-frequency g-modes, while $\delta$~Scuti stars host pulsations in low radial order (i.e., high-frequency p-modes). But the latter are also known to have low-frequency g-modes and mixed modes, which are driven by a combination of the $\kappa$-mechanism operating within the He II ionization zone \citep{Pamyatnykh2000} and the turbulent pressure in the hydrogen ionization layer \citep{Antoci_2014,Xiong_2016}. These stars form the $\delta$~Scuti instability strip in the HR diagram, which is located around the intersection of the classical instability strip and the main sequence \citep{dscuti_strip}.

High-mass hot stars (O, B, A) have convective cores and radiative envelopes, while low-mass solar-like stars (F, G, K) have radiative cores and convective envelopes. Accordingly, the magnetic fields observed in hot stars are generally fossil fields \citep{Borra1982}, i.e., dipolar stable fields that are remnants from the star's formation process, while the solar-like stars generally host dynamo fields. The position of $\delta$~Scuti stars between these two regimes makes them very important to study, as they can help us understand the transition between the two types of stellar structures and magnetic fields.

Previous studies have established that $\sim$ 10\% of O, B, and A stars have strong fossil magnetic fields \citep{mimes,BOB2015,LIFE2017}, with strengths of the order of 300-30,000 G \citep{Shultz2019_B_strength}. Among the $\delta$~Scuti stars, 13 magnetic stars have been confirmed to date \citep{neinerFirstDiscoveryMagnetic2015,Neiner2017, Zwintz2020, Keegan2021,Keegan2025}. But the overall incidence rate of magnetic $\delta$~Scuti stars remains unknown. The goal of this work is to determine this incidence rate and see if it is the same as the incidence rate of fossil magnetic fields in O, B, and A stars.

To obtain a statistically significant incidence rate of magnetic $\delta$~Scuti stars, a large sample of such stars is needed.
In order to directly detect magnetic fields on the surface of a star, we need spectropolarimetric observations. 
However, obtaining spectropolarimetric observations of such a large number of targets is a difficult task, especially because these targets span large ranges of magnitude and sky position. 
Fortunately, variable stars have been the targets of space missions that provide time series data. Such missions include that of the Convection, Rotation, and planetary Transits
(CoRoT) satellite (\citealt{Baglin2006Corot, Auvergne2009Corot}), the \textit{Kepler} mission \citep{Borucki2010Kepler}, and the Transiting Exoplanet Survey Satellite (TESS; \citealt{TESS2015}). 
Therefore, we could use an indirect method of detecting magnetism in variable stars based on their time series data. 
If a rotating star has a magnetic field, it can produce structures at the stellar surface (in temperature and/or chemical composition), resulting in a rotationally modulated light curve \citep{Mobster1}. 
In this work, we used the time series data from the CoRoT space mission to identify potentially magnetic stars among the $\delta$~Scuti stars observed by the satellite by looking for rotational modulation.

\section{Observations}\label{sec:obs}
Magnetic $\delta$~Scuti stars have been searched for among the targets of \textit{Kepler} and TESS missions using spectroscopic and spectropolarimetric observations \citep{thomson-paressantSearchMagneticScuti2023, Keegan2025} but not among the CoRoT targets because at the time of the CoRoT data release, $\delta$~Scuti stars were not expected to have magnetic fields. However, the sample of $\delta$~Scuti stars in the CoRoT database is very large and suitable for this work, as all of the variables in the CoRoT database have been identified using the CoRoT variability classifier \citep{eric_dscuti_2017}. We used the $\delta$~Scuti samples from this classification in this work.
This sample consists of 1757 objects classified as $\delta$~Scuti stars with a probability higher than 80\%. Although they have spectral-type and magnitude information in the CoRoT EXODAT database,\footnote{\url{https://cesam.lam.fr/exodat/}} we collected astrophysical parameters from \textit{Gaia} DR3 combined with the correction described in \citet{Fremat2024} for the effective temperature of hot stars in order to have more reliable information. We found that 68 of the targets have a spectral type of G or K, which are not known to host $\delta$~Scuti stars, so we removed them from our analysis.

\begin{figure}
  \resizebox{\hsize}{!}{\includegraphics{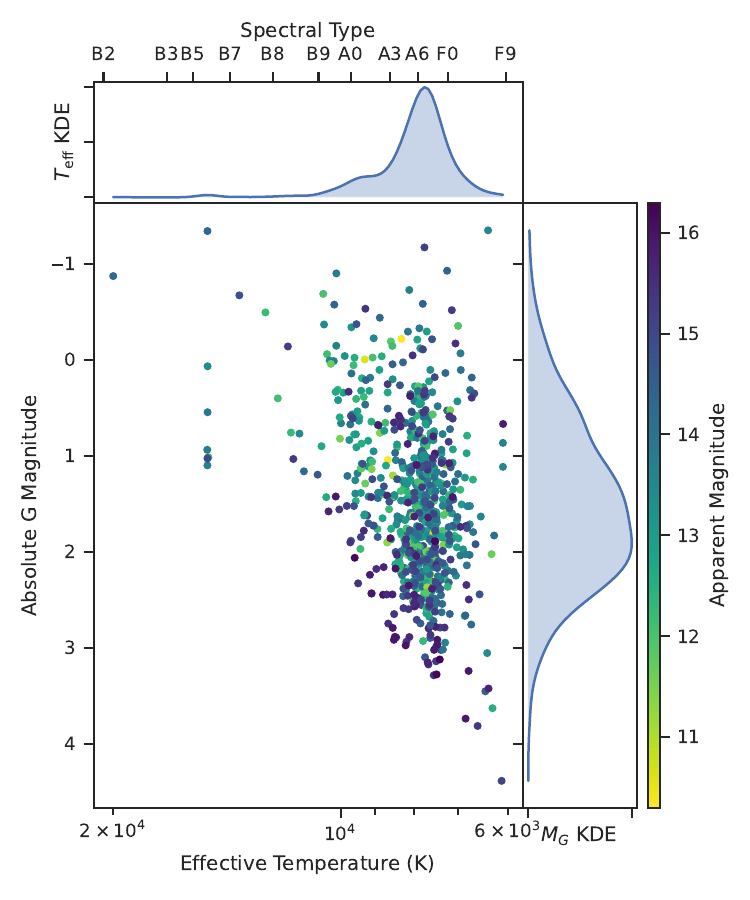}}
  \caption{Hertzsprung-Russell diagram showing a subset of the $\delta$~Scuti stars in the CoRoT catalog that have an absolute magnitude and effective temperature in \textit{Gaia} DR3. In the side plots of each axis, a kernel density estimation (KDE) of the corresponding variable is shown. This is an approximation of the underlying probability density function that generated the data. However, unlike a histogram, which bins and counts the observations, a KDE plot smooths the observations with a Gaussian kernel, producing a continuous density estimate.}
  \label{fig:hrd_all}
\end{figure}

In cases where effective temperature, magnitude, parallax, or extinction data were missing in \textit{Gaia} DR3, the CoRoT EXODAT database was used to obtain the spectral types, which were obtained using spectral energy distribution analysis \citep{exodat}. Conversion between the effective temperature and spectral type was done using the catalog of \citet{Mamajek2022_data}. The absolute magnitude was taken directly from the \textit{Gaia} row \texttt{mg\_gspphot} for the cooler stars and calculated using apparent magnitude, parallax, and extinction for the hotter stars.
Many of the targets had some of these parameters missing. The 780 targets for which we could obtain the absolute magnitude and effective temperature are shown in Fig. \ref{fig:hrd_all} as an HR diagram.

The CoRoT $\delta$~Scuti targets were observed using the ``faint star'' channel of the CoRoT 27 cm telescope. These stars have magnitudes between 10 and 16, and they were observed between February 2007 and November 2012.
The total of 1953 light curves, including multiple observations of some targets, were downloaded from the IAS CoRoT public archive's  ``\textit{faint stars - exo}'' database.\footnote{\url{http://idoc-corot.ias.u-psud.fr/}} These light curves were generated with a 512-second cadence and a baseline of 120-150 days (except some short light curves with a roughly 30-day baseline). 

All of these 1953 light curves are CoRoT N2 legacy data (version 4). The files labeled with the prefix \texttt{EN2\_STAR\_CHR} or \texttt{EN2\_STAR\_MON} respectively contain three ``pseudo-color'' light curves and a single monochromatic light curve. For both types of files, we used the monochromatic light curve from the \texttt{WHITEFLUXSYS} array and the time stamps from \texttt{DATEBARTT} array.

From the FITS file, \texttt{SYSTEMATIC} extension was used, which contained all the corrections from the CoRoT pipeline. These corrections include correction from aliasing, offsets, backgrounds, the jitter of the satellite, the change of the temperature set point, the loss of long-term efficiency, the jumps, replacement of the invalid and missing data using the inpainting method \citep{pires2015}, and residual systematics skews in the whole set of light curves of the run. Details of the CoRoT data formats and corrections are available in the CoRoT Legacy Book \citep{corot_legacy_2016}.

\section{Data analysis}
\subsection{Periodogram and noise level}
\begin{figure}
  \resizebox{\hsize}{!}{\includegraphics{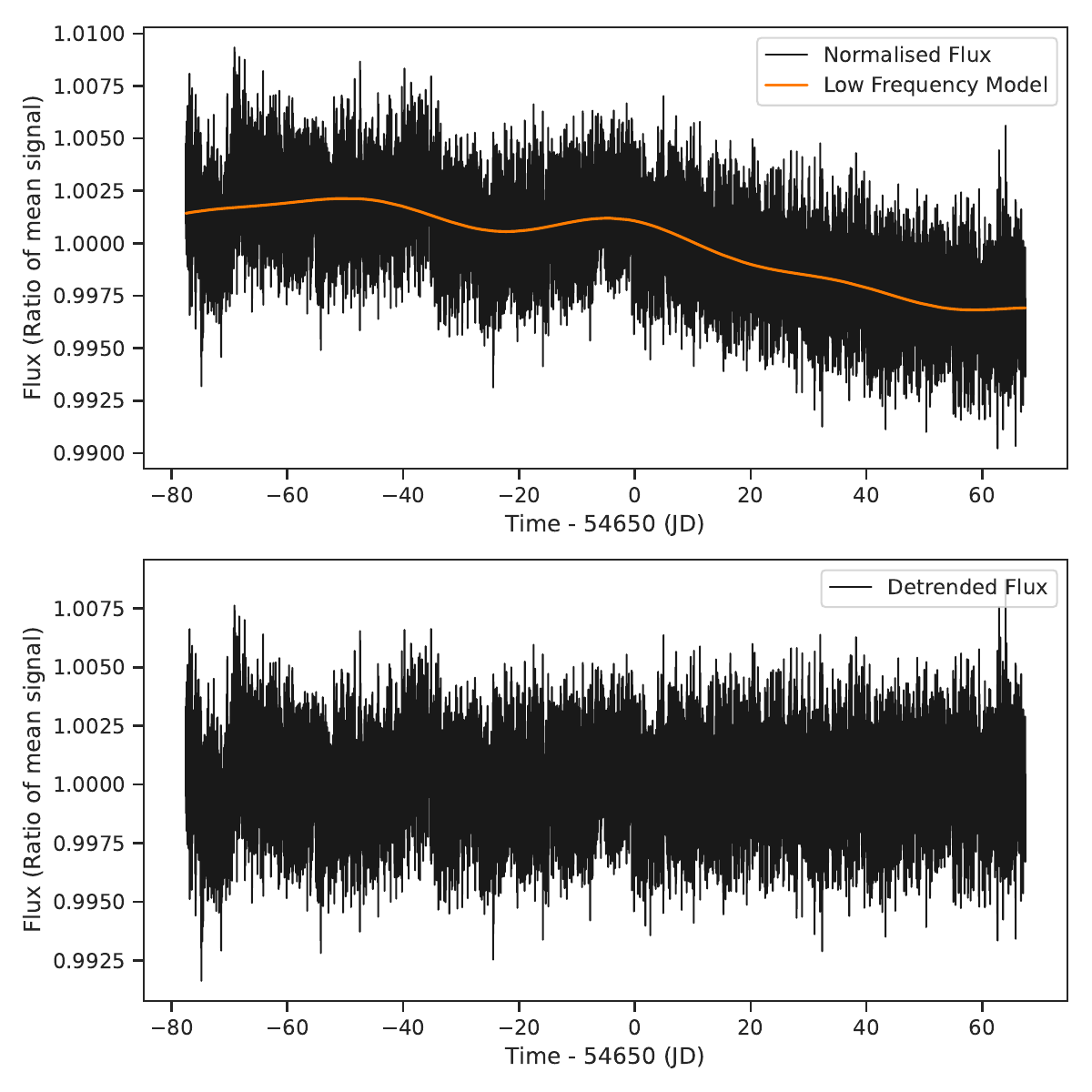}}
  \caption{Example light curve with a modeled long-term trend (top) and the detrended light curve (bottom). Both light curves have been normalized to have a unit mean signal.}
  \label{fig:lc_trend}
\end{figure}

Since the absolute value of the flux is not necessary for this analysis, the light curves were normalized by their median flux values.
Some of the light curves showed long-term trends and sudden jumps in the flux values. These are more likely to be instrumental effects than physical phenomena of the targets. We removed these long-term trends by modeling and removing the dominant frequencies lower than 0.025 $d^{-1}$ since a large majority of A/F stars rotate faster than this \citep{Royer2007}. 
We used the \texttt{LombScargle.model} method from the \textsc{Astropy} library to model and remove these frequencies using the prewhitening method described in Sect. \ref{sec:rot}.
An example of such a long-term trend and detrended light curve is shown in Fig. \ref{fig:lc_trend}.
Since the gaps in the data were filled with the inpainting method (see Sect. \ref{sec:obs}), there is basically no window except the frequency corresponding to the $512 s$ cadence, which is much higher than the rotation and $\delta$~Scuti pulsation frequencies. 

Some of the light curves contained a few outliers, which could be removed by using a $\sigma$ clipping method. However, removal of the outliers did not significantly change the power spectrum of the light curves, so we did not remove them.

To inspect the frequency spectra, periodograms were produced for each light curve. 
The primary tool we used to produce the periodograms was a modified generalized Lomb-Scargle periodogram \citep{Zechmeister2009,Press1992,VanderPlas2012,VanderPlas2015, Lomb, Scargle} employing the \textsc{timeseries.LombScargle} package of \textsc{Astropy} \citep{astropy2013, astropy2018}.

To determine if a peak in the periodogram is significant, it is necessary to define a metric. The \textsc{Astropy} library provides a method to calculate the false alarm probability (FAP) of a peak in the LS periodogram. The FAP is the probability that a peak of a given height could occur by chance in the noise of the data.
It is calculated by comparing the height of the peak to the noise level of the periodogram. The \textsc{Astropy} library does this by estimating a constant noise level for the periodogram and then calculating the FAP based on that noise level.
However, this method could not be applied directly to the light curves in our sample, as the noise level in our sample is frequency dependent, especially around very low frequencies, which is where the rotational modulation is expected to be found.
To solve this problem, we used a method to estimate the local noise level of the periodogram at each frequency. The noise level was estimated by calculating a running average of the periodogram of the light curve before detrending. To make the noise grid smoother, a Gaussian kernel with a standard deviation of 0.1 $d^{-1}$ was used to calculate the running average. The value of the standard deviation of the kernel was determined using a trial-and-error method.
In addition, to better estimate the rise of noise at low frequencies, where the running average function suffers from boundary effects, the spectra were reflected to the negative frequency range to remove the boundary effects at 0 $d^{-1}$. 
This method allowed for a more accurate estimation of the noise level at each frequency. The typical noise level found in the periodograms is $\sim 0.5\%$ of the mean signal in the $f > 3 d^{-1}$ range, and it is higher in the lower frequency range.

\subsection{Searching for rotational modulation} \label{sec:rot}
A rotational modulation has a frequency corresponding to the rotation of the star (fundamental frequency), usually with one to several of its harmonics. The amplitudes of the harmonics are expected to be decreasing with increasing order, and beyond the second harmonics, they usually have relatively low amplitudes. Therefore, we chose to look for the fundamental rotation frequency and its first and second harmonics.
The rotation frequency is expected to be under 3 $d^{-1}$ for $\delta$~Scuti stars because frequencies higher than 3 $d^{-1}$ would be supercritical. So we first searched for all the peaks under 7.5 $d^{-1}$, as the rotation frequency under 3.25 $d^{-1}$ with its first harmonic (and a rotation frequency under 2.5 $d^{-1}$ with its second harmonic) would be found in this range. The peaks were detected with a prewhitening method.

\paragraph{Prewhitening method.} The prewhitening method is a technique used to detect significant peaks in a periodogram by iteratively removing the effects of previously detected peaks. The method works as follows:
\begin{enumerate}
        \item Calculate the Lomb-Scargle periodogram of the light curve in the frequency range of interest.
        \item Identify the highest peak in the periodogram.
        \item Fit a sinusoidal function to the data with the frequency of the highest peak using \texttt{LombScargle.model} method from \textsc{Astropy}.
        \item Subtract the fit sinusoidal function from the original light curve to remove its effect.
        \item Repeat steps 1-4 until a predefined threshold is reached.
\end{enumerate}

The threshold for the prewhitening method was set to detect the 50 highest peaks.
The significance of a peak was determined by its signal-to-noise ratio (S/N).
In asteroseismology, an S/N threshold of 3.6 or 4 is most commonly used \citep{Breger1993}, but we set a lower threshold of 3 to be more inclusive in the search for magnetic candidates and also considering that the `noise' level may be artificially increased due to astrophysical signals in the vicinity of the identified peak.
It should be noted that this S/N is only that of the single frequencies. However, what we were actually looking for was rotational modulation: signals that have one or more harmonics. The presence of such harmonics would decrease the FAP of rotational modulation. 
With the frequency and amplitude of the peaks, we also kept track of the S/N of the peaks, as this would allow us to set a higher threshold later if needed.

Once all the significant peaks were found under 7.5 $d^{-1}$, for each peak ($f_0$), we checked if there were peaks at the first harmonic ($2 f_0$) and the second harmonic ($3 f_0$). When looking for peaks at the position of the harmonics, a frequency uncertainty of $\delta f = T_{lc}^{-1}$ was used around the expected frequency of the harmonic, with $T_{lc}$ being the total time span of the light curve.
If any of these harmonics were found, the fundamental frequency was considered a candidate for rotational modulation. 

While in most of the candidates only one peak was detected to have first or second harmonics, in some cases there were multiple peaks that had harmonics. To select the correct frequency of rotation among these candidate frequencies, several criteria were combined:
\begin{itemize}
        \item Frequency range: The fundamental frequency of rotation was expected to be in the range of 0.1 $d^{-1}$ to 3 $d^{-1}$ for $\delta$~Scuti stars.
        \item Correlation: When there were multiple peaks that had harmonics, some presumably fundamental frequencies were themselves harmonics of other peaks. In such cases, we selected the original fundamental frequency that was not a harmonic of any other peak.
        \item S/N: For each set of fundamental frequency and its harmonics, we checked the S/N of the peak that selected this set as a candidate and called it the "candidate S/N". This meant the following:
              \begin{itemize}
                      \item For a set containing a fundamental frequency ($f_0$) and first harmonic ($h_1$): \\ Candidate S/N = min(S/N($f_0$), S/N($h_1$)).
                      \item For a set containing a fundamental frequency ($f_0$) and second harmonic ($h_2$): \\ Candidate S/N = min(S/N($f_0$), S/N($h_2$)).
                      \item For a set containing a fundamental frequency ($f_0$), first harmonic ($h_1$), and second harmonic ($h_2$): \\ Candidate S/N = max(min(S/N($f_0$), S/N($h_1$)), min(S/N($f_0$), S/N($h_2$))).
              \end{itemize}
              The set with the higher candidate S/N was selected as the rotational frequency.
          
        \item Phase-folded light curve: For the cases of multiple candidates for rotational frequency, the light curve was folded at each of the candidate frequencies. After plotting the phase-folded light curve with a running average to smooth it out, a weak modulation of the light curve was revealed because the light curves still contained high-frequency pulsations. The frequency of the signal that seemed most consistent with rotation was selected as the rotational frequency.
    
        \item Rotational splitting: In the presence of rotation, the degeneracy of pulsation modes is lifted, leading to a splitting of the mode. Each non-radial ($l \ge 1$) mode in the pulsation spectrum was split into $2l + 1$ components, with azimuthal orders ($m$) ranging from $-l$ to $l$. For example, modes with $l = 1$ would be split into three frequencies corresponding to $m = -1, 0, 1$. The difference between these split frequencies is proportional to $m f_{rot}$, where $f_{rot}$ is the internal rotational frequency in the region of the mode's origin. As the $l=1$ modes are most visible, we looked at the whole range of pulsation frequencies and searched for the number of pairs of frequencies separated by a candidate rotation frequency ($\Delta f = f_0$), assuming $f_0 = f_{rot}$. The candidate rotational frequency with a higher number of such splittings present was considered more likely to be the true rotational frequency. But this is not an absolute criterion for the candidacy of the rotational frequency, as the pulsations can be influenced by the interior of the star, where the rotation is not the same as the surface rotation, and more rapidly rotating stars (faster than 50\% of Keplerian break up) have more complicated frequency spectra. Moreover, since $\delta$~Scuti stars generally have a high density of modes, some modes can be found to be separated by the candidate rotation frequency simply by chance. In this work, we did not analyze the splittings any further and treated their count as an optional criterion.
    
\end{itemize}

\subsection{Type of magnetic field} \label{sec:magtype}
If the rotational modulation is caused by a fossil magnetic field, all of its characteristics, such as amplitude and phase, are expected to be stable over time.
In contrast, if the rotational modulation is caused by a dynamo magnetic field, the amplitudes of the frequency peaks in the periodogram are expected to change over time.
To identify this, we divided the time series into several smaller windows of 40 days with 5 days of separation between each window. 
Then we calculated the periodogram in each window and looked for the amplitude of the candidate frequency in each periodogram. 
If the change of the amplitude of the candidate frequency was well above the local noise level (S/N > 3), then we considered the frequency to have a variable amplitude and the associated magnetic field to be caused by the dynamo. Conversely, if the change in the amplitude was not well above the local noise level, then we considered the frequency to have a stable amplitude and the associated magnetic field to be a fossil field.

\section{Results and discussion}
\subsection{Occurrence rate of magnetism in \texorpdfstring{$\delta$~Scuti}{delta Scuti} stars}
After the initial search for rotational modulation, we found 172 candidate light curves for rotational modulation belonging to 162 unique stars. After this automated search, we conducted a visual inspection of the 172 light curves using the criteria described in the previous section. After this inspection, we were able to discard 15 light curves where the fundamental frequency of rotation and the detected harmonics had a lower S/N and thus were not very distinguishable from the noise. The remaining 157 light curves belonging to 147 targets were considered to have rotational modulation. 
Although decreasing the S/N threshold to lower than three increases the number of magnetic candidates, most of these additional candidates are detected because of frequencies (fundamental frequency or harmonics) that are not very distinguishable from noise. A threshold of three for the S/N produced a low number of possible false detections, which we discarded. This way, we could ensure that most of the actual magnetic candidates were selected and that most of the false detections were removed.

\begin{figure}
  \resizebox{\hsize}{!}{\includegraphics{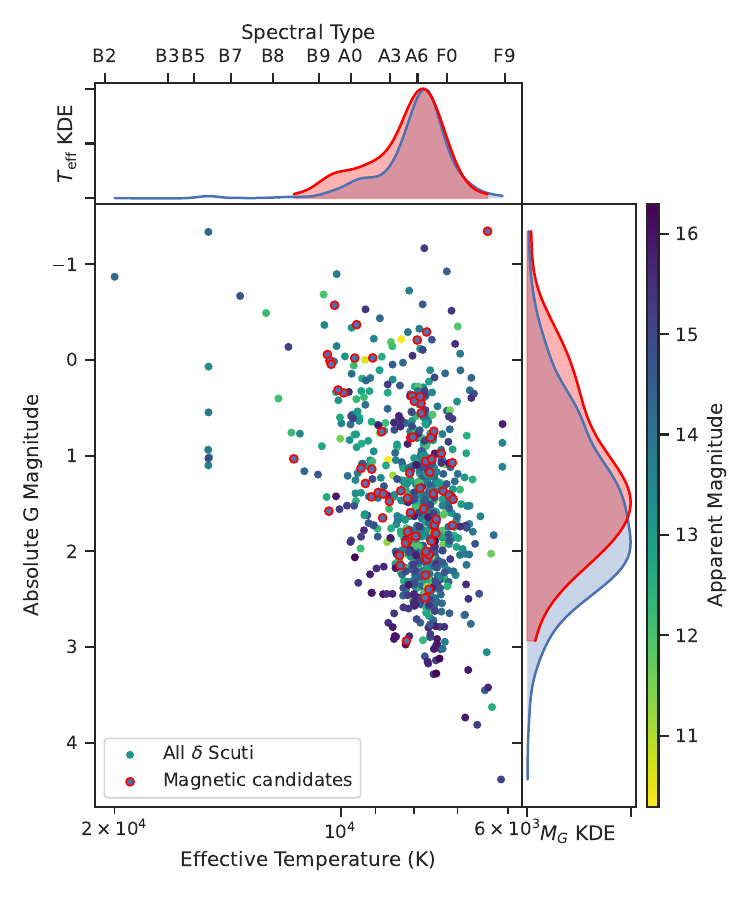}}
  \caption{Hertzsprung-Russell diagram showing the magnetic candidates among the $\delta$~Scuti stars in the CoRoT catalog. Shown are only 68 targets for which \textit{Gaia} parameters were found. The kernel density estimation (KDE) of $M_G$ and $T_{eff}$ shows that the distribution of magnetic candidates is similar to the distribution of all $\delta$~Scuti targets, except for a slight excess at higher temperature. However, the absolute magnitude distribution of the magnetic candidates is shifted slightly to lower values.}
  \label{fig:hrd_all_det}
\end{figure}

These 147 magnetic candidates constituted an 8.7\% incidence rate of magnetic candidate $\delta$~Scuti stars among the CoRoT $\delta$~Scutis. 
However, if we were to select only O, B, and A-type stars, the incidence rate of magnetic candidates among them would rise to 11.4\%. 
An HR diagram of these magnetic candidates is shown in Fig. \ref{fig:hrd_all_det}.

To have a selection of the candidates with a higher S/N, we created a subgroup with the ``high S/N'' candidates, which have S/N > 3.6. There are 83 high S/N candidates.
With this ``high S/N selection,'' the incidence rate of magnetic $\delta$~Scuti candidates falls to 4.9\% -- about half of the incidence rate with a ``low S/N selection.'' Among O, B, and A-type stars only, the high S/N incidence rate of magnetic candidates becomes 6.2\%.

\subsection{Rotational variability}
\begin{figure}
  \resizebox{\hsize}{!}{\includegraphics{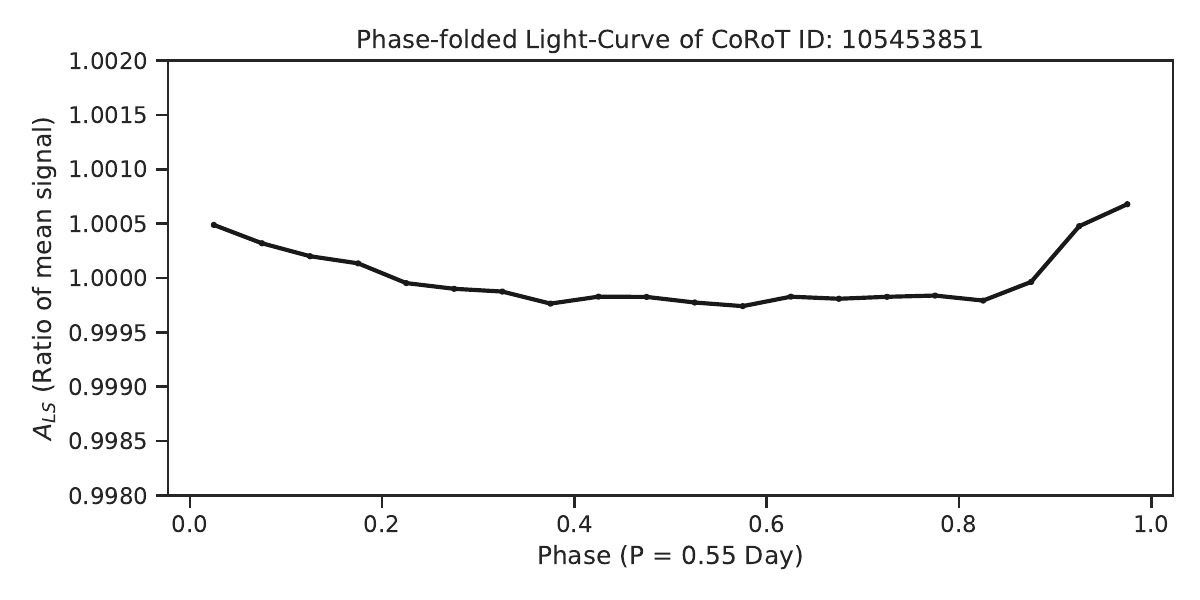}}
  \caption{Light curve of a magnetic candidate phase-folded at the period of the rotation. The plotted line is a running average, which is necessary to reveal the rotational modulation.}
  \label{fig:lc_phase}
\end{figure}

While selecting the rotational frequency of the candidates, the phase-folded light curve at the candidate rotational frequency was checked. As the light curve contained all the high- and low-frequency pulsation modes, a running average was plotted to smooth it out. This revealed the rotational modulation of that light curve (Fig. \ref{fig:lc_phase}). Such plots were used to select the correct rotational frequency in cases where multiple frequencies with harmonics were detected.

While the algorithm to search for rotational modulation looked for both the first and second harmonics of the fundamental frequency, we found that 114 candidates had only the first harmonic, 22 candidates had only the second harmonic, and 11 candidates had both the first and second harmonics. But this is the count of only the peaks that were detected in the periodogram using the S/N criteria mentioned in the previous section. In some cases, there might be more harmonics present that were not detected due to a low S/N. In addition, the initial search for the peaks was done until 7.5 $d^{-1}$, so if some harmonics were present at higher frequencies, they would not have been detected.

\begin{figure}
  \resizebox{\hsize}{!}{\includegraphics{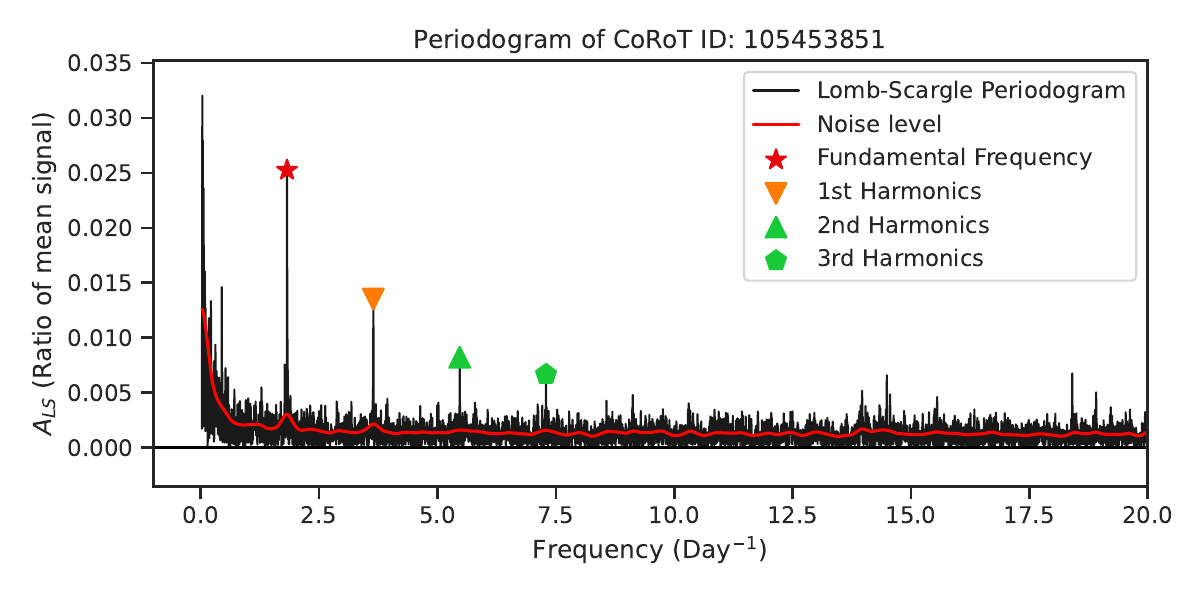}}
      \caption{Periodogram of a magnetic candidate where the harmonics of increasing order have decreasing amplitudes.}
  \label{fig:rot_good}
\end{figure}

\begin{figure}
  \resizebox{\hsize}{!}{\includegraphics{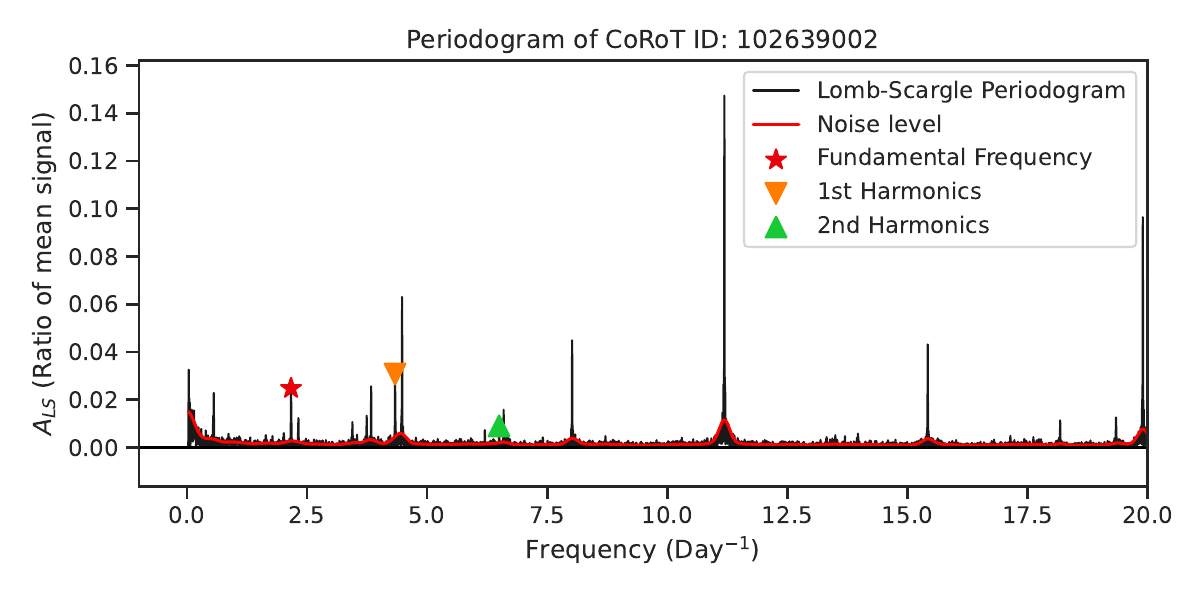}}
      \caption{Periodogram of a magnetic candidate where the first harmonic has a higher amplitude than the fundamental frequency.}
  \label{fig:rot_bad}
\end{figure}

The simplest rotational modulation pattern is characterized by a peak at the fundamental frequency and several harmonics of that frequency, with the fundamental frequency having the highest amplitude and the harmonics having decreasing amplitudes (Fig. \ref{fig:rot_good}). In our sample, 92 candidates had this pattern with the detected harmonics.
However, it is also not uncommon for the amplitude of the first harmonic to be higher than that of the rotation frequency. The presence of two diametrically opposite spots can cause such a modulation. We found that 55 candidates did not follow the decreasing amplitude pattern (Fig. \ref{fig:rot_bad}).

\begin{figure}
  \resizebox{\hsize}{!}{\includegraphics{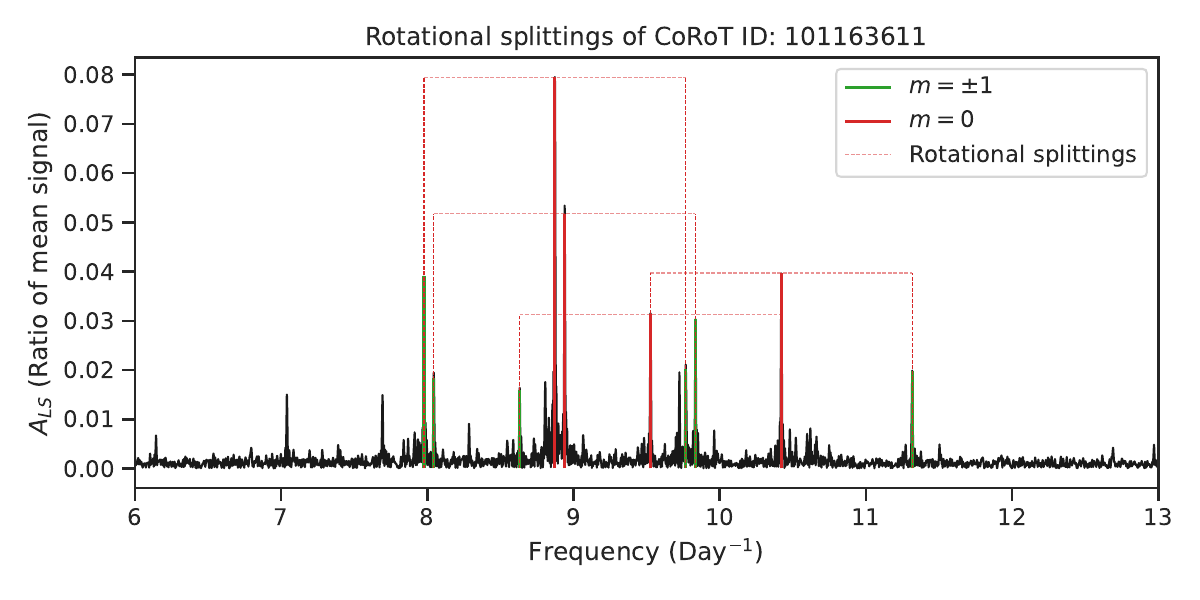}}
  \caption{Periodogram of a target where four triplets of frequencies are found. Each triplet has three modes separated by the candidate rotational frequency. The central mode of each triplet is indicated as $m=0$, and the other two as $m=\pm 1$. No actual mode identification was done in this work.}
    \label{fig:periodogram_spl}
\end{figure}

\begin{figure}
  \resizebox{\hsize}{!}{\includegraphics{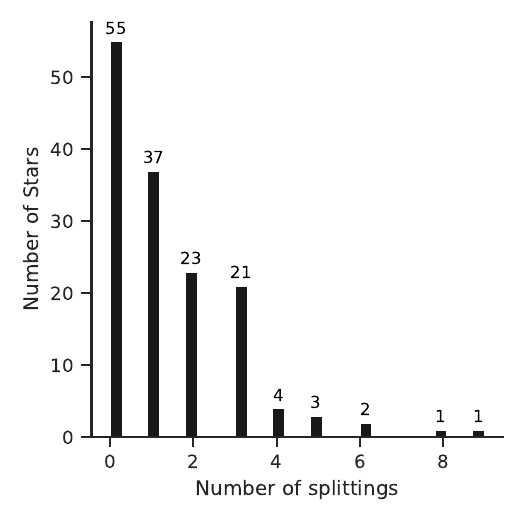}}
  \caption{Distribution of the candidates by the number of splittings of the pulsation modes at the candidate rotational frequency.}
    \label{fig:rot_splitting}
\end{figure}

The rotational candidates had between zero and nine p-mode pulsation frequencies, which were split with $\Delta f = f_0$. Their distribution is shown in Fig. \ref{fig:rot_splitting}. In the cases where there were multiple candidates for the rotational frequency, we selected the candidate with the highest number of splittings as the better candidate for the rotational frequency. The periodogram of a target with a high number of splittings is shown as an example in Fig. \ref{fig:periodogram_spl}.

\subsection{\texorpdfstring{$\delta$~Scuti--$\gamma$~Dor hybrids}{delta Scuti--gamma Dor hybrids}}
$\delta$~Scuti pulsations are p-mode pulsations and are expected to be in the frequency range of 5 $d^{-1}$ to 80 $d^{-1}$. However, in the sample, while all the candidates had $\delta$~Scuti pulsations, some candidates also had pulsations at frequencies lower than 5 $d^{-1}$.
Pulsations at these frequencies are usually high radial order g-mode pulsations of the $\gamma$~Dor type. However, peaks in this range can also originate from nonlinear coupling of p-modes. Additionally, $\delta$~Scuti stars can also have low radial order g-modes.
We found 90 rotational modulation candidates that had pulsations at both $\delta$~Scuti and $\gamma$~Dor frequency ranges. These are possible $\delta$~Scuti--$\gamma$~Dor hybrid stars with a magnetic field, which are particularly interesting candidates for magneto-asteroseismology. Fifty-seven candidates had pulsations only in the $\delta$~Scuti frequency range, indicating pure $\delta$~Scuti stars with a magnetic field.

\subsection{Binary star}
From inspection of the light curves and periodograms, we found three targets whose light curves were contaminated by a transiting companion, suggesting an eclipsing binary companion around each, and one target with a possibly contaminated light curve but without any transit signal. Their CoRoT IDs are 101351899, 605144067, 110848138, and 102700458. In their contaminated light curves, it is not possible to find a rotational modulation that indicates a magnetic field. Therefore, these targets have been excluded from the list of magnetic candidates.

The light curve of the candidate 101351899 showed clear periodic dips in the flux values, indicating eclipses at a period of 11 days. The periodogram showed a peak at 0.09 $d^{-1}$, which corresponds to the period of the eclipses and several ($\sim 40$) harmonics of that peak. A phase-folded light curve of this star is shown in Fig. \ref{fig:bin_1}, where it shows a clear sign of binary eclipse rather than rotational modulation. This target was also identified as an eclipsing binary in \citet{binary_1} with a transit period of 10.99 days.

The light curve of the candidate 110848138 showed clear periodic dips in the flux values, indicating eclipses at a period of 3.5 days. The periodogram showed a peak at 0.28 c/d, which corresponds to the period of the eclipses and several harmonics of that peak. The phase-folded light curve of this star shows a clear sign of binary eclipse.

The candidate 102700458 showed many high-amplitude peaks with several harmonics, which is not typical for $\delta$~Scuti pulsation or rotational modulation. The phase-folded light curve revealed a binary eclipse signal.

For the candidate 605144067, the light curve and the periodogram did not show the regular pattern of $\delta$~Scuti pulsations. This suggests that this target was either possibly misclassified as a $\delta$~Scuti star or the light of the $\delta$~Scuti star was polluted by a non-$\delta$~Scuti companion, which would be a visual binary. However, no eclipsing signal was found in the light curve.

Except for these four binary candidates, we found $\delta$~Scuti pulsations in the periodogram of all the other rotational modulation candidates. This indicates a very low occurrence rate of possible false detection for $\delta$~Scuti stars in the initial list of the CoRoT $\delta$~Scuti sample.

\subsection{Fossil and dynamo magnetic fields}
\begin{figure}
  \resizebox{\hsize}{!}{\includegraphics{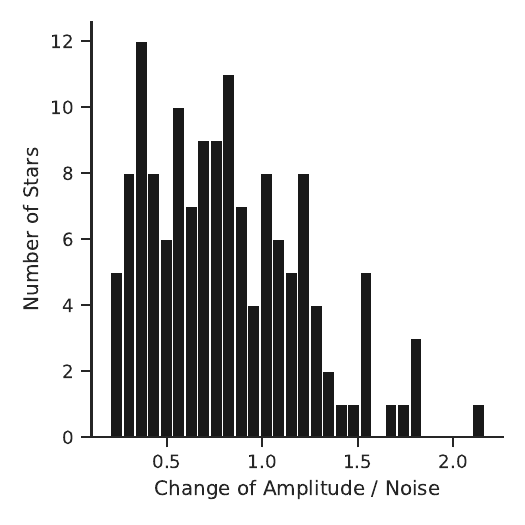}}
  \caption{Distribution of the S/N of the change of amplitude over time. This plot contains the data of 141 targets. The remaining six targets have short light curves, so we did not compute their change of amplitude.}
  \label{fig:dA_dt_dist}
\end{figure}

\begin{figure}
  \resizebox{\hsize}{!}{\includegraphics{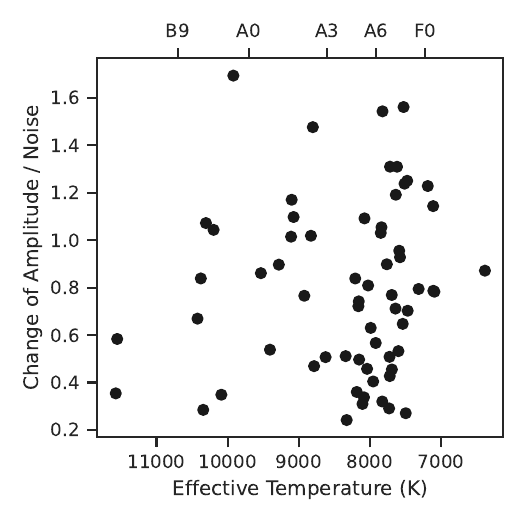}}
  \caption{Change of amplitude over time versus effective temperature. This plot includes the 69 targets with an effective temperature available in \textit{Gaia} DR3.}
  \label{fig:dA_dt_Teff}
\end{figure}

Our sample consists of stars from spectral type B2 to F9, and the magnetic candidates have spectral type B2 to F8. The hotter OBA-type stars are expected to have a fossil field, while the cooler stars are expected to have dynamo fields. Among our 147 magnetic candidates, we tried to characterize the type of their magnetic field using the algorithm described in \ref{sec:magtype}. We found that all of the 147 candidates have a stable amplitude of the rotational frequency, i.e., the change in the amplitude of the rotational frequency is very low compared to the local noise level. This parameter is plotted as a distribution in Fig. \ref{fig:dA_dt_dist}. Among the 147 targets, 69 of them have an effective temperature in \textit{Gaia} DR3. Figure \ref{fig:dA_dt_Teff} shows the change of amplitude over noise against the effective temperature, but we did not observe any trend against the temperature. If there were targets with a significant change of amplitude over noise for the magnetic peak ( > 3), we would expect these targets to have lower effective temperatures. However, the absence of such targets indicates that all the magnetic candidates host a rather stable field on the timescale of the CoRoT light curve. 

\subsection{Magnetic candidates for future spectropolarimetric observations}
The candidates for magnetic $\delta$~Scuti stars from this work need to be confirmed with spectropolarimetric observations. Such observations will allow for measurement of the magnetic field strength and its geometry at the surface of the star. To select suitable candidates for spectropolarimetric observations, one needs to consider the brightness of the star, its position in the sky, spectral type or temperature, period of rotation, and the projected rotational velocity ($v \sin i$). 
The period of rotation is important because it determines the time resolution required for the observations and the phase coverage of the rotation. In addition, the required S/N in the spectropolarimetric observations, and therefore the required exposure time, decreases with decreasing effective temperature, as the square root of the number of spectral lines that can be used in the analysis increases. The amplitude of the spectropolarimetric signal is proportional to the slope of the line profile, and therefore it decreases as $(v \sin i)^{-2}$, so the required exposure time increases as $(v \sin i)^2$. With the candidate rotation period selected in this work, we calculated the projected rotational velocity ($v \sin i$) of the candidates using the formula 
\begin{equation}
    v \sin i =  \frac{2 \pi R}{P_{rot}} <\sin i>
\label{eq:vsini}.
\end{equation}
The terms are as follows:
\begin{itemize}
    \item The radius of the star is $R$. The radii of the stars were estimated from their spectral types using the catalog of \citet{Mamajek2022_data}. 
    \item The average value of $\sin i$ is $<\sin i>$, which is statistically equal to $\frac{2}{\pi}$. 
    \item The rotation period $P_{rot}$ is the candidate rotational period and is calculated by $P_{rot} = f_{rot}^{-1}$, where $f_{rot}$ is the candidate rotational frequency.
\end{itemize}

\begin{figure}
  \resizebox{\hsize}{!}{\includegraphics{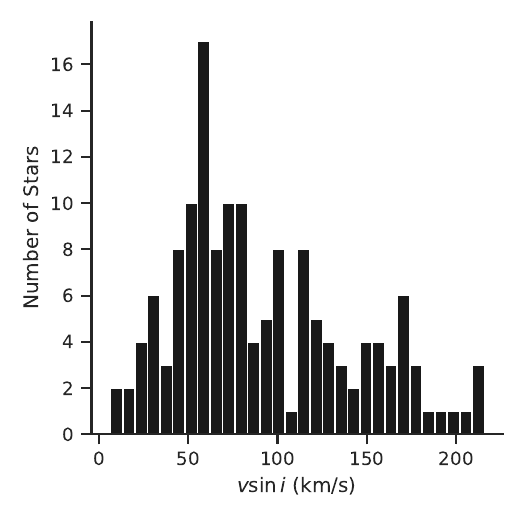}}
  \caption{Distribution of the $v \sin i$ of the magnetic candidtes}
  \label{fig:vsini_dist}
\end{figure}

A distribution of the $v \sin i$ estimates is shown in Fig. \ref{fig:vsini_dist}. With these parameters, one can select the candidates for spectropolarimetric observations based on the specific spectropolarimeter and telescope used for the observations.

\section{Conclusions}
In this work, we have presented a method to search for magnetic $\delta$~Scuti candidates in the CoRoT light curves and created a Python tool to facilitate that search. Using this tool, we analyzed 1953 light curves of 1689 $\delta$~Scuti stars in the CoRoT sample. Among them, we found 147 good candidates for magnetic $\delta$~Scuti stars. We classified these candidates according to their S/N, presence of rotational splitting, variability type, and the number of harmonics detected.

We conclude that the incidence rate of magnetic $\delta$~Scuti candidates in the CoRoT sample is 8.7\%  for the full sample and 11.4\% if we consider only the OBA stars. This is close to the incidence rate of fossil magnetic fields in OBA stars, which is around 10\%. The latter incidence rate was measured by spectropolarimetric studies. Our work is solely based on photometric light curves, and it yielded a similar incidence rate for $\delta$~Scuti stars. However, these incidence rates should be considered an upper limit, as close binaries or chance alignment of lower frequency signals could possibly produce a rotational modulation signal. 
This is especially important for the hotter stars in our sample because a large fraction of stars with a higher mass have binary companions, so there may be cases where, for instance, a rotationally variable B-type star is paired with a cooler $\delta$~Scuti companion, which we would not be able to distinguish in the CoRoT data.
Furthermore, it is hard to distinguish between signals from ellipsoidal variables and rotational modulation because they produce a similar pattern in the frequency spectrum. To determine our candidates, we selected those from our sample that have phase-folded light curves more consistent with a rotational modulation signal (e.g., Fig. \ref{fig:lc_phase} is consistent with rotational modulation), but there are cases where it is ambiguous as to whether the variation is rotational or due to binary-induced ellipsoidal variation. \textit{Gaia} DR3 provides the renormalized unit weight error (RUWE) associated with each target.
The RUWE is expected to be around 1.0 for sources where the single-star model provides a good fit to the astrometric observations. A value significantly greater than 1.0 could indicate that the source is non-single or otherwise problematic for the astrometric solution. Among the 147 magnetic candidates, 124 have RUWE < 1.5, indicating that a greater proportion of our candidates are less likely to be binaries. Further investigation with spectropolarimetric surveys of these targets will reveal the binary nature and/or confirm the existence of a magnetic field.

According to this work, based on CoRoT light curves, all 147 candidates have stable, possibly fossil, magnetic fields. This result does not follow our expectation, as cooler stars in this sample should instead have dynamo fields. 
The actual types of the magnetic fields need to be verified with complementary methods, such as spectropolarimetry. In addition, dynamo fields are hard to detect using light curves only, and some $\delta$~Scuti stars from this sample may also host dynamo fields that have remained invisible in this study. 

With the European Space Agency's PLAnetary Transits and Oscillations of Stars (PLATO; \citealt{plato2014, plato2025}) mission, it will be possible to obtain light curves with higher quality and less noise. However, PLATO's long observation fields do not include the magnetic candidates in this work, and the selection of the shorter observation fields has not been finalized yet. If some of the latter PLATO fields include the magnetic candidates in this work from the CoRoT field, then it will be possible to use the tools developed for this work to analyze the PLATO light curves of these candidates and obtain additional information about the magnetic candidates. 

Finally, we have prepared the necessary data and parameters for the candidates to select the suitable targets for spectropolarimetric observations. These observations will allow us to measure the magnetic field strength and its geometry at the surface of the star, which will help us confirm the presence of magnetic fields in $\delta$~Scuti stars and better understand their nature.
However, due to the generally low brightness of the CoRoT targets, only a few of these candidates are suitable for spectropolarimetry with current instruments, such as Echelle SpectroPolarimetric Device for the Observation of Stars (ESPaDOnS) at the Canada-France-Hawaii Telescope. Better instruments on larger telescopes will be needed in the future for further analysis of these stars, such as the Visible and Infrared spectroPolarimetER (VIPER) for the Very Large Telescope at ESO.

\section*{Data availability}
The full version of Table~\ref{tab:mag_dsct} is only available in electronic form at the CDS via anonymous ftp to \url{cdsarc.u-strasbg.fr} (130.79.128.5) or via \url{http://cdsweb.u-strasbg.fr/cgi-bin/qcat?J/A+A/}.

\begin{acknowledgements}
We thank Eric Michel for providing the list of $\delta$~Scuti targets in the CoRoT database. 
This work is funded/co-funded by the European Union (ERC, MAGNIFY, Project 101126182). Views and opinions expressed are, however, those of the authors only and do not necessarily reflect those of the European Union or the European Research Council. Neither the European Union nor the granting authority can be held responsible for them.
This work is based on observations from the Convection, Rotation and Planetary Transits (CoRoT) space telescope.
The CoRoT space mission, launched on 2006 December 27, was developed and is operated by the CNES, with participation of the Science Programs of ESA, ESA's RSSD, Austria, Belgium, Brazil, Germany, and Spain.
This work has made use of data from the European Space Agency
(ESA) mission \textit{Gaia} (\href{https://www.cosmos.esa.int/gaia}{https://www.cosmos.esa.int/gaia}), processed by
the \textit{Gaia} Data Processing and Analysis Consortium (DPAC, \href{https://www.cosmos.esa.int/web/gaia/dpac/consortium}{https://www.
cosmos.esa.int/web/gaia/dpac/consortium}). 
This research has made use of the SIMBAD database operated at CDS, Strasbourg (France), and NASA’s Astrophysics Data System (ADS).
We also thank the anonymous referee for their useful comments.
\end{acknowledgements}

\bibliographystyle{aa}
\bibliography{Biblio}

\begin{appendix}
\section{Binary star}
\begin{figure}[ht]
  \resizebox{\hsize}{!}{\includegraphics{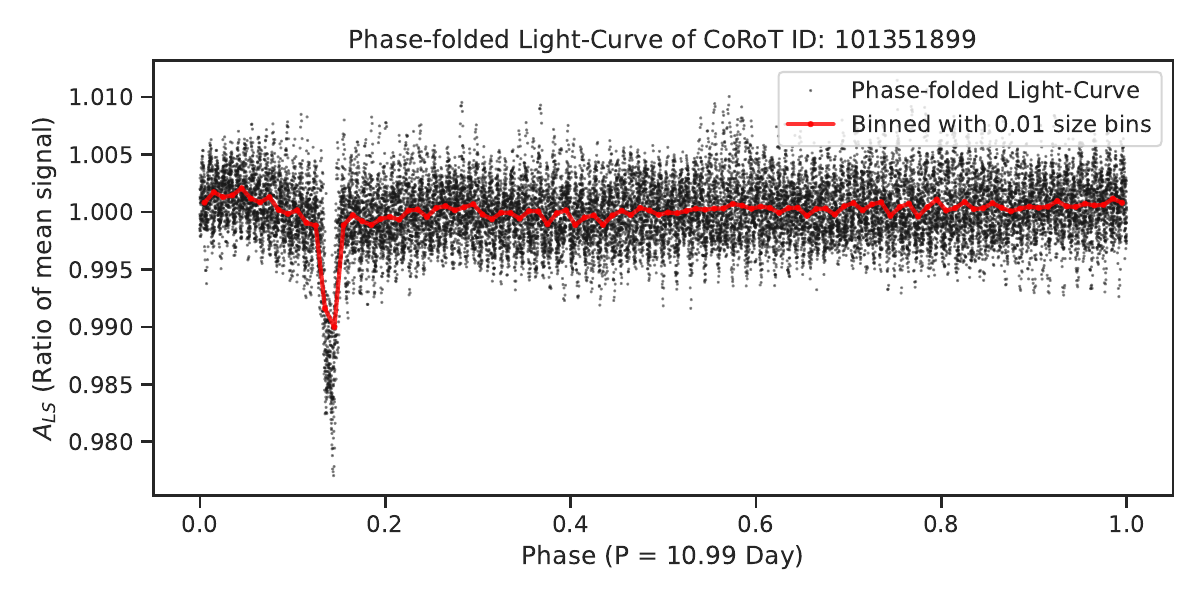}}
  \caption{Light curve of a binary candidate phase-folded at the period of the binary eclipse. This signature of a binary eclipse is distinctively different from that of a rotational modulation. }
  \label{fig:bin_1}
\end{figure}

Of the four binary candidates in our sample,three showed binary eclipse signals in their phase-folded light curve (e.g., Fig. \ref{fig:bin_1}).

\section{Magnetic candidates}
\begin{table}[ht]
        \caption{Detected magnetic $\delta$~Scuti candidates with their rotation frequencies, $f_{rot}$ (extract). }
        \label{tab:mag_dsct}
    \centering
    \begin{tabular}{llrrr}
                \hline\hline
                CoRoT ID & \textit{Gaia} source ID & RA (deg) & Dec (deg) & $f_{rot}$ (d$^{-1}$) \\
                \hline
100445224 & 4288627786833193344 & 290.634036 & 1.599151 & 0.79 \\
100560095 & 4264551750688074752 & 290.804539 & 1.356114 & 1.55 \\
101083805 & 4287832672447259008 & 291.584888 & 1.410531 & 0.80 \\
101163611 & 4288700049686196352 & 291.710043 & 2.009709 & 0.89 \\
101246854 & 4287847515854670464 & 291.840786 & 1.712671 & 2.93 \\
101370668 & 4287501719443254272 & 292.061554 & 0.399719 & 2.70 \\
101392446 & 4287775257322817152 & 292.102166 & 1.352298 & 1.19 \\
101488855 & 4263426293133377536 & 292.267059 & 0.176719 & 2.09 \\
102291417 & 3318183327916490624 & 92.314858 & 4.518866 & 1.07 \\
102308606 & 3318687423935532288 & 92.530815 & 5.446701 & 1.04 \\
102308661 & 3318612347904962048 & 92.531626 & 5.050650 & 1.71 \\
102324359 & 3318596027032147456 & 92.730685 & 4.943212 & 3.02 \\
102337356 & 3317839253793518592 & 92.877050 & 4.784809 & 1.32 \\
102345902 & 3318643026857726336 & 92.973362 & 5.260428 & 1.01 \\
102354673 & 3318643477830615680 & 93.074872 & 5.305299 & 0.90 \\
\hline 
\end{tabular}
\tablefoot{The sky positions are given in the equinox 2000 reference frame. The full table is only available in electronic form at the CDS.}.
\end{table}

\end{appendix}
\end{document}